\documentclass[aps,prl,twocolumn,showpacs]{revtex4}
\usepackage{amssymb}
\usepackage{amsfonts}
\usepackage{amsmath}
\usepackage{graphicx}
\begin{document}


\title{Evidence for Half-Metallicity in n-Type HgCr$_2$Se$_4$}

\author{Tong Guan, Chaojing Lin, Chongli Yang}
\affiliation{Beijing National Laboratory for Condensed Matter Physics, Institute of Physics, Chinese Academy of Sciences, Beijing 100190, China}

\author{Youguo Shi}
\email{ygshi@iphy.ac.cn}
\affiliation{Beijing National Laboratory for Condensed Matter Physics, Institute of Physics, Chinese Academy of Sciences, Beijing 100190, China}

\author{Cong~Ren}
\email{cong_ren@iphy.ac.cn}
\affiliation{Beijing National Laboratory for Condensed Matter Physics, Institute of Physics, Chinese Academy of Sciences, Beijing 100190, China}

\author{Yongqing~Li}
\email{yqli@iphy.ac.cn}
\affiliation{Beijing National Laboratory for Condensed Matter Physics, Institute of Physics, Chinese Academy of Sciences, Beijing 100190, China}

\author{Hongming Weng, Xi Dai and Zhong Fang}
\affiliation{Beijing National Laboratory for Condensed Matter Physics, Institute of Physics, Chinese Academy of Sciences, Beijing 100190, China, and Collaborative Innovation Center of Quantum Matter, Beijing 100084, China}

\author{Shishen Yan}
\affiliation{School of Physics, Shandong University, Jinan 250100, China}

\author{Peng Xiong}
\affiliation{Department of Physics, Florida State University, Tallahassee, Florida 32306, USA}

\begin{abstract}
 High quality HgCr$_2$Se$_4$ single crystals have been investigated by magnetization, electron transport and Andreev reflection spectroscopy. In the ferromagnetic ground state, the saturation magnetic moment of each unit cell corresponds to an integer number of electron spins (3\,$\mu_B$/Cr$^{3+}$), and the Hall effect measurements suggest n-type charge carriers. Spin polarizations as high as $97\%$ were obtained from fits of the differential conductance spectra of HgCr$_2$Se$_4$/Pb junctions with the modified Blonder-Tinkham-Klapwijk theory. The temperature and bias-voltage dependencies of the sub-gap conductance are consistent with recent theoretical calculations based on spin active scatterings at a superconductor/half metal interface. Our results suggest that n-HgCr$_2$Se$_4$ is a half metal, in agreement with theoretical calculations that also predict undoped HgCr$_2$Se$_4$ is a magnetic Weyl semimetal.

\end{abstract}


\maketitle



Chromium chalcogenides of the spinel group ACr$_2$X$_4$ (A=Hg, Cd, Zn, X=Se, S) have been studied as magnetic semiconductors or insulators for several decades~\cite{Baltzer65,Baltzer66,Goldstein78}. They have attracted much recent interest because the ferromagnetism and band structures are conducive to the emergence of Chern semimetallicity~\cite{XuG11}. Chern semimetals (i.e. magnetic Weyl metals) are a class of ferromagnetic materials in which band crossings are guaranteed to exist at certain points in the momentum space by topological protection~\cite{XuG11}. In addition to the fascinating Weyl fermion physics~\cite{WanXG11,Hosur13,WengHM15} that was recently observed in their nonmagnetic counterparts~\cite{LiuZK14a,LiuZK14b,XuSY15,LvBQ15,HuangXC15}, Chern semimetals can also host other interesting phenomena requiring broken time-reversal symmetry, such as the quantized anomalous Hall effect~\cite{YuR10,ChangCZ13}. As a candidate proposed for Chern semimetals, HgCr$_2$Se$_4$ can also be a unique spintronic material due to its $s$-band half-metallicity when doped to $n$-type~\cite{XuG11}.

Ferromagnetic order in ACr$_2$X$_4$ arises at low temperatures due to the superexchange interactions between the 3d electrons of Cr. The $s$-$d$ exchange interactions can produce a large spin splitting of the conduction bands [Fig.\,1(a)]. Depending on the magnitude of exchange splitting and the strength of spin-orbit coupling, these chalcospinels can be either magnetic semiconductors (e.g. CdCr$_2$Se$_4$~\cite{Lehmann67}) or Chern semimetals (e.g. HgCr$_2$Se$_4$~\cite{XuG11}). Regardless of whether the exchange splitting is strong enough for a band inversion, a common feature of these chromium chalcospinels is that there exists a sizable range of chemical potential for fully spin polarized $s$-band conductivity~\cite{XuG11,GuoSD12}; this makes them a unique type of half-metals. Half-metals are a class of magnetic materials in which the charge current is conducted by carriers of one spin orientation~\cite{Coey04,Katsnelson08}. In a half-metal, the Fermi level is located either inside an energy gap or in the localized states for one spin direction~\cite{note1}, whereas the electronic states are extended for the other. The complete spin polarization of itinerant carriers makes half-metals very attractive for spintronic applications, such as in high performance magnetic tunnel junctions~\cite{Bowen03} and  for efficient spin injection into semiconductors~\cite{Schmidt00}. Recent theories also suggest that half-metals, especially those with single bands and strong spin-orbit interactions, may host p-wave or topological superconductivity~\cite{Chung11,Duckheim11} when they are in proximity to s-wave superconductors~\cite{Keizer06,Eschrig08}. So far, however, only a handful of materials have been shown experimentally to be half-metallic ~\cite{JiY01,Parker02,Steeneken02,Braden03}. Moreover, charge carriers in these materials either have $d$-band characteristics (e.g. CrO$_2$~\cite{Korotin98}, (La,Sr)MnO$_3$~\cite{Bowen03,Coey99}, EuO~\cite{Schiller01}) or are holes in $p$-orbitals (e.g. GaMnAs)~\cite{Dietl14}. Definitive experimental demonstration of $s$-band half-metals remains elusive, and no direct measurement of the electron spin polarization has been reported in this spinel class of materials.

In this letter, we report that measurements of magnetization, electron transport and Andreev reflection spectra have been carried out on high quality $n$-type HgCr$_2$Se$_4$ single crystals. Exchange splitting induced metal-insulator transition is manifested as 8 orders of magnitude change in longitudinal resistivity upon ferromagnetic ordering as well as colossal magnetoresistance near $T_C$. In the ferromagnetic state, the saturation magnetization  is found to correspond to 3.0\,$\mu_B$/Cr$^{3+}$. The zero-bias conductances of Pb/HgCr$_2$Se$_4$ Andreev junctions are nearly completely suppressed. These results coherently point to $n$-type HgCr$_2$Se$_4$ being a half-metal with carriers in the conduction band.

\begin{figure}
\includegraphics[width=8.5 cm]{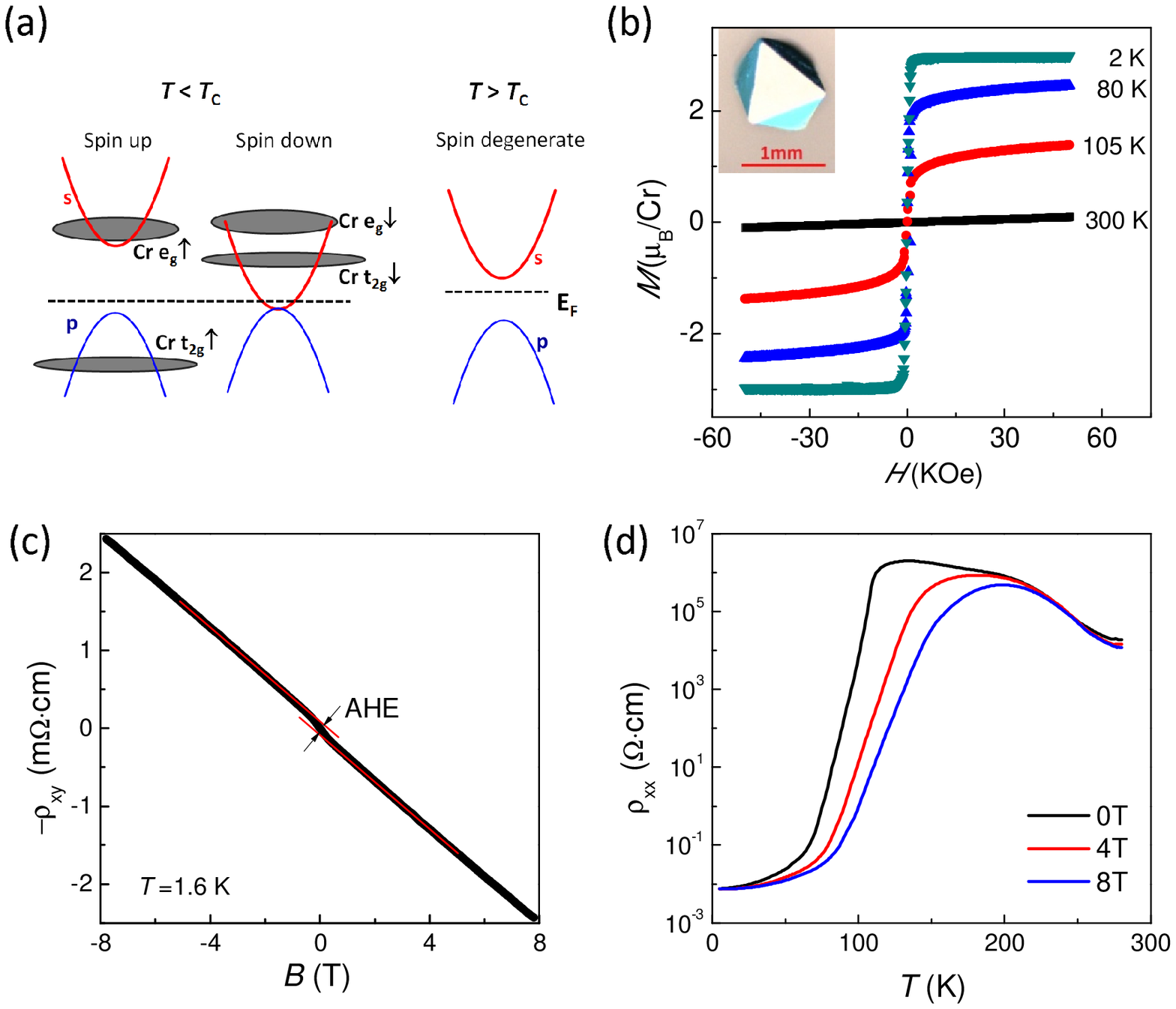}
\caption{(color online) (a) Schematic band diagrams for HgCr$_2$Se$_4$ in the ferromagnetic (FM) (left) and paramagnetic (right) states based on the first principles calculations reported in Refs.~\cite{XuG11,GuoSD12}.  (b) Magnetization curves at several temperatures from $T=$2\,K to 300\,K. The inset shows an optical micrograph for a typical HgCr$_2$Se$_4$ crystal. (c) Hall resistivity $\rho_\mathrm{xy}$ curve taken at $T=1.6$\,K. The small curvature near the zero-field is due to the anomalous Hall effect (AHE). (d) Temperature dependence of longitudinal resistivity $\rho_\mathrm{xx}$ at magnetic fields $\mu_0 H=$0, 4, 8\,T.}
\end{figure}

Single crystals of HgCr$_2$Se$_4$ were grown with chemical vapor transport~\cite{Supplem}.  The single crystalline samples usually are octahedral shaped and $\sim 1$\,mm in size [Fig.\,1(b) inset]. As shown in Fig.\,1(b), HgCr$_2$Se$_4$ is a soft ferromagnet below the Curie temperature ($T_C$=105.5\,K), and no hysteresis was clearly resolved.
The saturation magnetization $M_s$ corresponds to $3.00\pm0.05$\,$\mu_B$ per Cr$^{3+}$ ion.  In contrast, all previous works reported non-integer values of the magnetic moment (2.7-2.9\,$\mu_B$/Cr$^{3+}$)~\cite{Baltzer65,Baltzer66,Auslender88}. The observation of integer values of magnetic moment per unit cell satisfies a necessary condition and is an important indicator for the material being a half-metal~\cite{Coey04}. This is consistent with the results of the first principles calculations~\cite{XuG11,GuoSD12}. As illustrated in the band diagram in Fig.\,1(a), the measured magnetization corresponds to the fully occupied spin up t$_{2g}$ orbital of Cr. At low temperatures, Hall resistivity $\rho_\mathrm{xy}$ has a linear dependence on magnetic field, except at very low fields where magnetization is not saturated and the small curvature is cause by the anomalous Hall effect [Fig.\,1(c)]. The linearity in the normal Hall resistance for a wide range of magnetic fields enabled us to extract an electron density of about $2\times10^{18}$\,cm$^{-3}$ at $T<70$\,K. The $n$-type samples used for the Andreev reflection measurements to be presented below were taken from the same batch, and the Hall effect measurements consistently yielded carrier densities on the same order of magnitude and electron mobilities of the order of $10^2$\,cm$^2/$V$\cdot$s at $T=2$\,K.

In addition to the ideal magnetization value, the high quality of our $n$-HgCr$_2$Se$_4$ samples is also shown by the strong temperature dependence of the longitudinal resistivity $\rho_\mathrm{xx}$.  As depicted in Fig.\,1(d), $\rho_\mathrm{xx}$ increases by more than \emph{eight} orders of magnitude as $T$ is increased from 10\,K to temperatures above the Curie temperature. This is in contrast to only up to 3-4 orders of magnitude change in $\rho_\mathrm{xx}$ obtained in pervious works~\cite{Samokhalov81,Solin08}. The temperature dependence of $\rho_\mathrm{xx}$ can be attributed to a metal-insulator transition driven by the ferromagnet-paramagnet transition. In the paramagnetic phase, the chemical potential is located inside the bulk band gap. Below $T_C$, the s-d exchange interaction spin-splits the conduction band, and gradually raises the chemical potential above the conduction band minimum. The $\rho_\mathrm{xx}$ data shown in Fig.\,1(d) span from $\sim10^{-2}$\,$\Omega\cdot$cm (metallic) to $\sim10^6$\,$\Omega\cdot$cm (very insulating). Such a drastic variation in resistivity is comparable to the best values of magnetic semiconductors or other materials exhibiting colossal magnetoresistance (CMR) effect~\cite{Coey99}. The data in Fig.\,1(d) also indicate that the HgCr$_2$Se$_4$ sample has very large magnetoresistance (MR) around the phase transition region. The MR, defined here as $\mathrm{MR}=\rho_\mathrm{xx}(0)/\rho_\mathrm{xx}(B)-1$, reaches $7\times10^4$ at $B=8$\,T and $T=110$\,K.

We used the Andreev reflection spectroscopy to directly probe the electron spin polarization in HgCr$_2$Se$_4$~\cite{BTK82,Mazin01,Woods04}. It has been used to measure the spin polarization in many ferromagnetic materials~\cite{Upadhyay98}, including several candidates for half-metals, such as (La,Sr)MnO$_3$~\cite{Soulen98}, CrO$_2$~\cite{JiY01,Parker02}, (Ga,Mn)As~\cite{Braden03}, EuS~\cite{RenC07}. This method is based on the fact that the Andreev reflection probability at the interface between a superconductor and a ferromagnet is partially or completely suppressed by spin-imbalanced density of states at the Fermi level in the ferromagnet. The spin polarization can be extracted by fitting the conductance spectra to the modified BTK theory~\cite{BTK82,Mazin01,Woods04}. The measurement can be implemented either with point contact geometry~\cite{Soulen98} or with planar junction geometry~\cite{Parker02}. Both geometries have been proven effective in revealing the half-metallicity in CrO$_2$, a classic half-metal that has also been confirmed with other methods, such as Meservey-Tedrow experiment~\cite{Parker02}.

\begin{figure}
\includegraphics[width=8.5 cm]{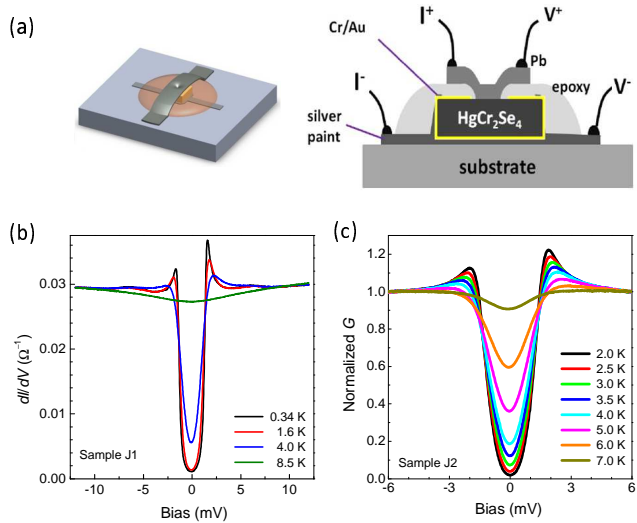}
\caption{(color online). (a) Sketch of the HgCr$_2$Se$_4$/Pb planar junctions used for Andreev reflection spectroscopy. (b) Differential conductance $G(V)=dI/dV$ of the sample J1 plotted as function of dc bias. (c) Normalized differential conductance $G(V)/G_N(V)$ spectra for sample J2.  Here, the bias $V$ is defined as the dc voltage of the Pb electrode relative to the Cr/Au electrode at the bottom of the HgCr$_2$Se$_4$ crystal, and $G_N(V)$ is normal state differential conductance recorded with a small field ($\mu_0 H=0.2$\,T) applied.}
\end{figure}

Fig.\,2(a) shows a sketch of the HgCr$_2$Se$_4$/Pb planar junctions used in this work.  A typical device was fabricated on a lustrous, triangle-shaped surface of a HgCr$_2$Se$_4$ single crystal [Fig.\,1(b), inset].
The differential conductance, defined as $G(V)=dI(V)/dV$, was measured as a function of the dc bias voltage $V$ for HgCr$_2$Se$_4$/Pb junctions using phase-sensitive lock-in detection. The amplitude of the ac modulation was kept sufficiently small in order to avoid heating and other spurious effects. Fig.\,2(b) shows the conductance curves of  sample J1 for temperatures from 0.34\,K to 8.5\,K. The Pb films used in this work has a superconducting transition temperature $T_c$=7.2\,K, nearly identical to the bulk $T_c$ value in literature.
The conductance data at 8.5\,K are close to the normal state values, $G_N(V)$, which were recorded when a perpendicular magnetic field $\mu_0 H=0.2$\,T was applied at corresponding temperatures.  For sample J1, $G_N$ is about 34\,$\Omega$, and it has a weak bias dependence.

Many other samples with junction resistances, $R_N$, in the range of 14-75\,$\Omega$ have also been studied. Although these values are low in comparison to $R_N$ of Andreev junctions on other low carrier density materials, an estimate of the junction area based on the point contact models~\cite{Yanson98} and these junction resistances yields values significantly smaller than the geometric junction areas. Possible explanations for this discrepancy are discussed in detail in Supporting Material~\cite{Supplem}. Fig.\,2(c) shows a set of normalized conductance ($G(V)/G_N(V)$) curves for a HgCr$_2$Se$_4$/Pb junction (sample J2) with $R_N\simeq 14$\,$\Omega$. The low junction resistance indicates a relatively weak barrier strength at the HgCr$_2$Se$_4$/Pb interface, thus offering the possibility of extraction of the electron spin polarization with the modified BTK theory. This is evidenced by the analysis of the conductance spectra later. These junction resistances are in contrast to those of the planar Andreev junctions of EuS, a magnetic semiconductor with much higher electron densities. In the latter, $R_N$ is on the order of 10\,k$\Omega$ and the $I$-$V$ trace is highly nonlinear due to the formation of a Schottky barrier~\cite{RenC05}. As shown in Ref.~\cite{Parker03}, very strong barrier strength could lead to the conductance spectra hardly discernible from the tunneling spectra between an unpolarized metal and a superconductor. The conductance spectra of our low-$R_N$ HgCr$_2$Se$_4$/Pb junctions have the following common characteristics: (1) The subgap conductance is significantly suppressed, as expected for a half-metal. (2) There exist a pair of small, but noticeable superconducting coherence peaks at the superconducting gap edges, i.e.\ $\mathrm{e}V=\pm\Delta$. The energy gap $\Delta$ is 1.4-1.5\,meV, close to the bulk value of Pb. (3) The heights of both coherence peaks are significantly lower than what is expected of a \textit{tunnel} junction, and the peak height at the positive bias (see Fig.\,2) is greater than that at the negative bias. Such an asymmetry has not been observed in the Andreev junctions of CrO$_2$ and EuS, despite that the barrier strength can be varied systematically from very weak to very strong. However, it has been observed in the Andreev reflection spectra of several strongly correlated electron systems, such as heavy fermion superconductor CeCoIn$_5$~\cite{Park08} and ferromagnetic semimetal EuB$_6$~\cite{ZhangXH08}. Further work is needed to confirm whether there is a link between the asymmetry and the electron correlation.

\begin{figure}
\includegraphics[width=8.5 cm]{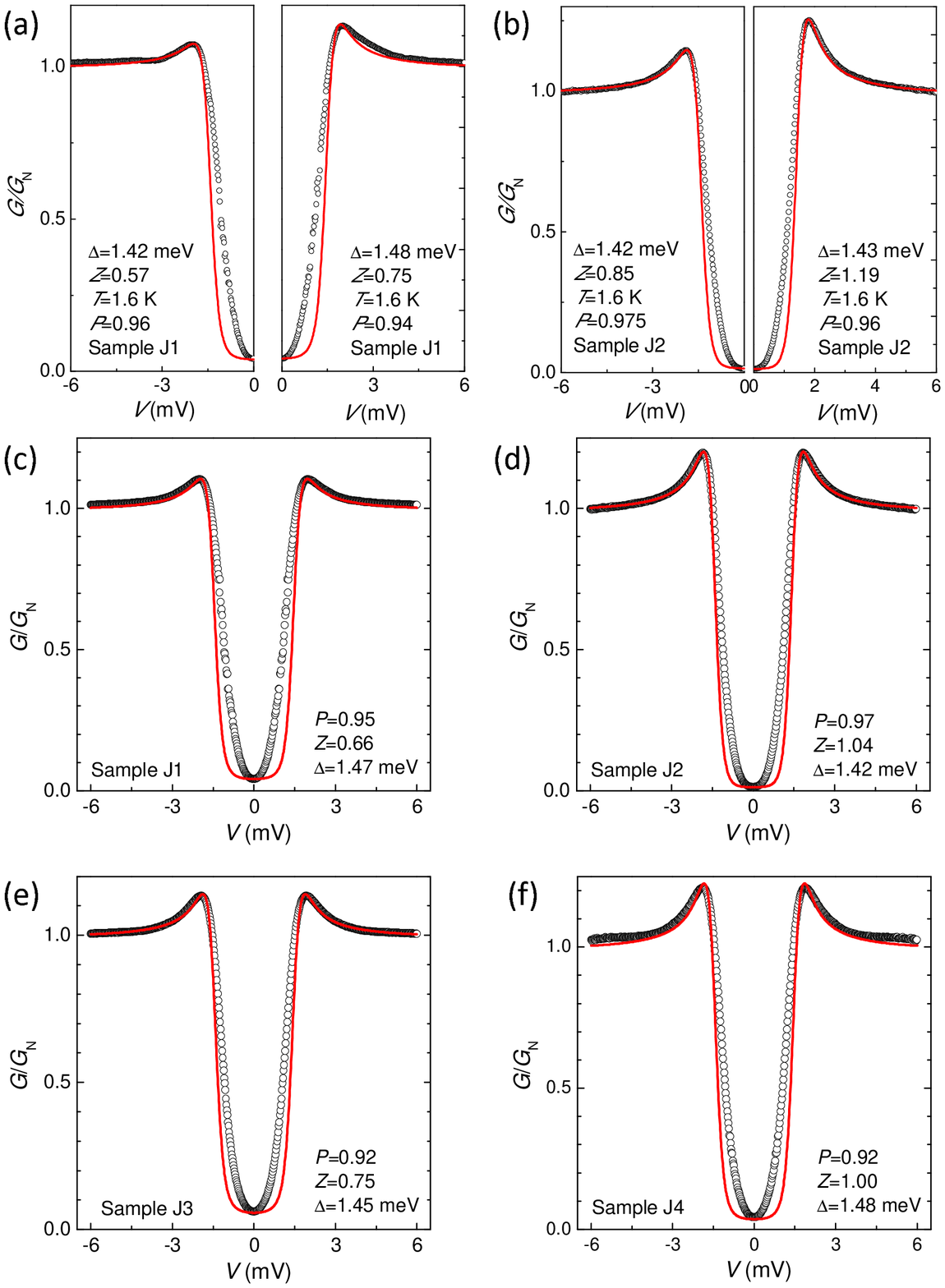}
\caption{(color online). Normalized conductance spectra (symbols) before symmetrization (Panels a-b, for samples J1 and J2) and after symmetrization (Panels c-f, for samples J1-J4).  Their fits to the modified BTK theory are shown in lines. The measurement temperature was 1.6\,K for all panels.}
\end{figure}

 The differential conductance data was analyzed with the modified BTK theory~\cite{Mazin01}, in which there are four fitting parameters: spin polarization $P$, barrier strength parameter $Z$, superconducting gap $\Delta$ and inelastic broadening parameter $\Gamma$.
  For the data in this work, we argue that $\Gamma$ can be approximated to be zero, for the following considerations: (1) In the spectra shown in Fig.\,2(b), both a superconducting gap of bulk Pb value and the Pb phonon features (around $\pm7$\,meV) are apparent, implying a clean HgCr$_2$Se$_4$/Pb interface with minimal lifetime effects. (2) The inelastic broadening energies extracted from Pb/I/Al, and especially Pb/I/CrO$_2$ junctions subjected to a similar surface treatment process~\cite{Parker02} is on the order of tens of $\mu$eV, two orders magnitude smaller than the gap energy. (3) Importantly, we have systematically examined the effects of varying $\Gamma$ on the BTK fits to our data: Best fits with increasing $\Gamma$ up to 0.5\,meV result in little change in the extracted $P$; although larger $\Gamma$ leads to decreased $P$, it is accompanied by deterioration of the fitting quality, especially near the coherence peaks.
   Figs.\,3(a) and 3(b) show the normalized conductance curves and the BTK fits with three adjustable parameters (i.e.\ $P$, $Z$ and $\Delta$)  for samples J1 and J2, respectively. The fits were carried out separately for the positive and negative halves of the conductance spectra, namely $G(V)$ at $V>0$ or $V<0$, and for each sample the extracted $P$ only differs up to a few percent for different polarities. This justifies the BTK fits with the symmetrized spectra (i.e.\, $G(V)\rightarrow [G(V)+G(-V)]/2$) depicted in Figs.\,3(c-f) for four junctions. The fits yield $P=95\%$, $97\%$, $92\%$ and $92\%$, and $Z$=0.66, 1.04, 0.75 and 1.00, for samples J1-J4, respectively.  There is no apparent trend of decreasing in $P$ with increasing $Z$, consistent with the previous work on planar CrO$_2$/Pb junctions~\cite{Parker02}.

The conductance spectra of the HgCr$_2$Se$_4$/Pb junctions and the extracted high degree of spin polarization from the modified BTK fits have some resemblance to those of CrO$_2$ planar junctions, in which spin polarizations of $90\%$ to $97\%$ were obtained for a wide range of barrier strengths ($Z=$ 0 to 2.7)~\cite{Parker03}. Nevertheless, it should be pointed out that a $100\%$ spin polarization (or correspondingly, completely suppressed subgap conductance) has \emph{never} been observed in any Andreev reflection experiment on CrO$_2$ or other half-metals despite an enormous amount of experimental efforts. Recent theoretical advances have rendered several mechanisms that can account for the small deviation from full spin polarization~\cite{Lofwander10}. These include various spin active scatterings at the superconductor/half-metal interface~\cite{Eschrig08,Beri09,Kupf09,Linder10,Kupf11,Wilken12}, the spin-orbit coupling in the superconductor~\cite{Duckheim11} and inelastic scatterings of quasiparticles in the half-metal~\cite{Beri09b}. These scatterings can relax the rule of spin conservation and lead to finite conductance at zero bias. Furthermore, the lifting of the particle-hole symmetry away from zero-bias brings a conductance correction of the form $\delta G\propto V^2$, if $V$ is not too large~\cite{Beri09,Kupf11}. Following these models, one would expect a \emph{quadratic} dependence of the junction conductance on the bias voltage. Indeed, this is borne out in the conductance spectra of HgCr$_2$Se$_4$/Pb junctions (Fig.\,3): As shown clearly in Fig.\,4(a), the low bias $G(V)$ of sample J2 deviates substantially from the BTK fit, but follows very well a quadratic function.

\begin{figure}
\includegraphics[width=8.5 cm]{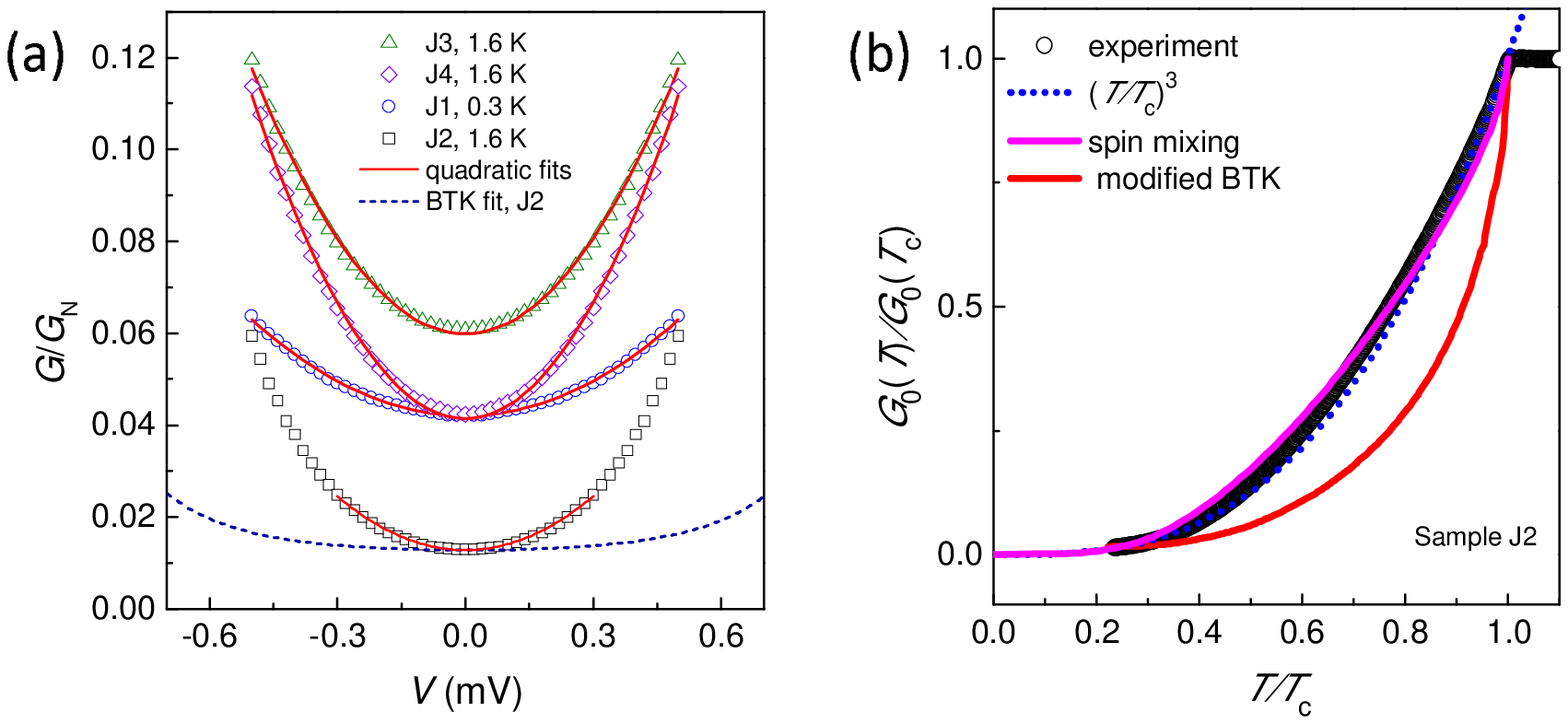}
\caption{(a) Normalized conductances at small bias voltages of HgCr$_2$Se$_4$/Pb junctions at low temperatures ($T=0.3$\,K for samples J1, and 1.6\,K for samples J2-J4, symbols). Each of them can be fitted to a quadratic function, i.e. $G(V)\sim G_0+ bV^2$ (solid lines). The fit of the conductance spectra of sample J2 to the modified BTK theory (dotted line) is also shown for comparison. (b) Temperature dependence of the zero-bias conductance $G_0$ (symbols), which roughly follows a cubic dependence on $T$ (dotted line). It is also close to the calculations of the spin active scattering model with spin singlet-triplet mixing angle $\vartheta=\pi/4$ (solid line)~\cite{Lofwander10}. The values derived from the modified BTK fit (solid line at the bottom) deviates strongly from the experimental curve. The data were taken from sample J2 in zero magnetic field.}
\end{figure}

Further evidence for the deviation from conventional BTK-type Andreev reflection comes from the temperature dependence of zero-bias conductance $G_\mathrm{0}$. As shown in Fig.\,4(b), $G_\mathrm{0}$ of sample J2 roughly follows a $T^3$ dependence: $G_\mathrm{0}/G_\mathrm{0,N}\simeq(T/T_C)^3$. In contrast, the modified BTK model predicts substantially smaller $G_\mathrm{0}$ values at intermediate temperatures. The excess of conductance at finite temperatures is qualitatively in good agreement with the calculation of L\"{o}fwander et al. with a model of spin active scatterings at a clean superconductor/half-metal interface for the case of singlet-triplet mixing angle $\vartheta=\pi/4$~\cite{Lofwander10}. An interesting outcome of their work is that $G(V)$ vanishes at the $T=0$ limit. As shown in Figs.\,2(b) and \,2(c), the zero-bias conductance $G_0(T)$ saturates at very small values as $T$ approaches zero. The low temperature $G_0$ appears to vary from sample to sample [Fig.\,4(a)], suggesting the interface quality is important. This may lend support to the disorder assisted spin active scatterings proposed in Ref.~\cite{Wilken12} for finite $G_0$ at $T=0$. The additional conductance induced by the spin active scatterings may be responsible for the underestimation of the spin polarizations in half-metals with the fits to the modified BTK theory reported in the literature~\cite{Soulen98,Lofwander10}. The values obtained from the modified BTK fits may only provide lower bounds for the spin polarization.

In summary, the high degree of spin polarization derived from fits to the modified BTK theory as well as the integer moment magnetization provide strong evidence for half-metallicity in n-HgCr$_2$Se$_4$. This is consistent with the first principle calculations that predicted the half-metallic and Chern semimetallic phases for the n-doped and undoped HgCr$_2$Se$_4$, respectively~\cite{XuG11}. Furthermore, the small but finite zero-bias conductance and its temperature dependence are compatible with recent theoretical models developed for spin-active scatterings, which have been proposed as a venue for realizing the exotic triplet paring with the superconducting proximity effects in half-metals~\cite{Eschrig08,Beri09,Chung11,Duckheim11}.

\begin{acknowledgements} \emph{Acknowledgements:} This work was supported by the National Basic Research Program (Projects No.\ 2015CB921102, No.\ 2012CB921703, No.\ 2011CBA00108, No.\ 2011CBA00110 and No.\ 2013CB921700), the National Science Foundation of China (Projects No.\ 91121003, No.\ 11374337 and No.\ 61425035), and the Chinese Academy of Sciences. P.X. acknowledges support from NSF grant DMR-1308613.
\end{acknowledgements}

\end{document}